\documentclass[10pt]{iopart}

\usepackage{graphicx}
\usepackage[english]{babel}
\usepackage{color}
\usepackage{multirow}

\newcommand{\eis}{EuIr$_2$Si$_2$}
\newcommand{\ers}{EuRh$_2$Si$_2$}
\newcommand{\ecs}{EuCo$_2$Si$_2$}

\newcommand{\enp}{EuNi$_2$P$_2$}

\begin{document}

\title[Charge, lattice and magnetism in EuIr$_2$Si$_2$]{Charge, lattice and magnetism across the valence crossover in EuIr$_2$Si$_2$ single crystals}

\author{Silvia Seiro$^{1,2}$, Yurii Prots$^1$, Kurt Kummer$^3$, Helge Rosner$^1$, Ra\'ul Cardoso Gil$^1$, Christoph Geibel$^1$}

\address{$^1$ Max Planck Institute for Chemical Physics of Solids, N\"{o}thnitzer Stra{\ss}e 40, D-01187 Dresden, Germany}
\address{$^2$ IFW-Dresden, Helmholtzstra{\ss}e 20, D-01069 Dresden, Germany}
\address{$^3$ European Synchrotron Radiation Facility (ESRF), 71 avenue des Martyrs, CS 40220, F-38043 Grenoble cedex 9, France}
\ead{s.seiro@ifw-dresden.de}

\date{\today}

\begin{abstract}
We present a detailed study of the temperature evolution of the crystal structure, specific heat, magnetic susceptibility and resistivity of single crystals of the paradigmatic valence-fluctuating compound \eis. A comparison to stable-valent isostructural compounds \ecs\ (with Eu$^{3+}$), and \ers, (with Eu$^{2+}$) reveals an anomalously large thermal expansion indicative of the lattice softening associated to valence fluctuations.  A marked broad peak at temperatures around 65-75\,K is observed in specific heat, susceptibility and the derivative of resistivity, as thermal energy becomes large enough to excite Eu into a divalent state, which localizes one f electron and increases scattering of conduction electrons. In addition, the intermediate valence at low temperatures manifests in a moderately renormalized electron mass, with enhanced values of the Sommerfeld coefficient in the specific heat and a Fermi-liquid-like  dependence of resistivity at low temperatures. The high residual magnetic susceptibility is mainly ascribed to a Van Vleck contribution.  Although the intermediate/fluctuating valence duality is to some extent represented in the interconfiguration fluctuation model commonly used to analyze data on valence-fluctuating systems, we show that this model cannot describe the different physical properties of \eis\ with a single set of parameters.  

\end{abstract}

\submitto{\JPCM}

\section{Introduction}

The valence of Eu can be influenced by its chemical environment~\cite{Segre82}, pressure~\cite{Wada99}, temperature~\cite{Bauminger73} or magnetic field~\cite{Matsuda08}. Not only 4f$^6$ (Eu$^{3+}$) and 4f$^7$ (Eu$^{2+}$) configurations can be realized, but also non-integer valence states~\cite{Bauminger73}. The Eu valence has dramatic consequences for the physical properties due to the different number of conduction electrons, volume, and magnetic behavior of Eu$^{2+}$  (large, pure-spin $J$=7/2 moment, ordering at low temperatures) and Eu$^{3+}$ ($J$=0 ground state with Van Vleck paramagnetism). In a first approximation, the non-integer valence can be interpreted as a fluctuating state between the two integer valence configurations that lie close in energy. Upon warming, the high-energy state, which in valence-fluctuating Eu systems is always the divalent one, becomes increasingly populated and the average valence shifts towards 2. Similarly, a magnetic field promotes the strongly magnetic divalent state, while pressure favors the non-magnetic trivalent configuration because of its much smaller volume. In M\"{o}ssbauer spectroscopy, a technique that probes nuclear transitions leaving electronic states unaffected~\cite{Kohn82},  an intermediate isomer shift between those of the integer Eu valence states is observed~\cite{Bauminger73}. In probes such as X-ray absorption, which result in strong changes of intermediate or final electronic states~\cite{Kohn82}, one observes two lines at the positions corresponding to the integral valences~\cite{Mitsuda00}. At low temperatures, this simple thermal excitation picture fails, notably in \eis~and \enp, as the valence measured in M\"{o}ssbauer experiments remains non-integer down to the lowest temperatures~\cite{Nagarajan85, Chevalier86}. This implies the presence of a significant hybridization between 4f and conduction electrons, which results at low temperatures in a quantum mechanical mixing of the two integer valence states. This hybridization leads e.g. to a sizeable mass renormalization of the conduction electrons, as attested by the large Sommerfeld coefficient extracted from specific heat measurements~\cite{Fisher95, Seiro11}. The optical response as a function of temperature also contradicts a simple thermal occupation scheme, and resembles remarkably that of Kondo lattice systems, where hybridization between localized and conduction electrons plays the dominant role~\cite{Guritanu12}.
In this work, we investigate the effect of valence fluctuations on the lattice, magnetic, and transport properties of the paradigmatic valence-fluctuating compound \eis\, by comparing with isostructural, stable-valent compounds \ecs~ and \ers, which have the same total number of electrons. We disentangle contributions from valence fluctuations at intermediate temperatures, and from an intermediate valence at low temperatures. We show that the interconfiguration fluctuation mode, which considers thermal excitation to the divalent state and the effect of hybridization on a very simple empirical level, provides only a limited description of the data.

\section{Experimental details}

Single crystals of Eu\textit{T}$_2$Si$_2$  were grown by a flux method (\textit{M}=Rh,Ir)~\cite{Seiro11} or by a Bridgman technique (\textit{M}=Co)~\cite{Seiro13} starting from high purity elements. Temperature-dependent synchrotron powder diffraction data on crushed single crystals were collected in transmission (Debye-Scherrer) geometry at the high-resolution beamline (ID22) of the European Syncrothron Radiation Facility (particle size $< 32$\,$\mu$m, capillary of 0.3\,mm diameter, $\lambda$ = 0.400737(6)\,\AA, $3^{\circ} \leq 2\theta \leq 52^{\circ}$, step $2\theta=0.002^{\circ}$, multi-channel detector, $10\,\rm{K}\leq \rm{T}\leq 293\,K$, liquid-helium cryostat). The Rietveld refinement was performed with the program package WinCSD~\cite{Akselrud14}. The valence of Eu was investigated by means of X-ray absorption spectroscopy (XAS) performed at the former ID16 beamline of the European Synchrotron Radiation Facility (now ID20). The spectra were collected in partial fluorescence yield using an X-ray emission spectrometer equipped with a Ge (333)  analyzer crystal and set to the Eu $\rm L\alpha_1$ emission line at 5.843 keV ($\Delta E = 0.86\,$eV). In this way, the intrinsic core-hole lifetime broadening of the spectra is significantly reduced which allows to detect and quantify spectral changes more easily than in conventional, total fluorescence yield XAS spectroscopy~\cite{Hamalainen91,deGroot02,Kummer11}. Magnetization data were acquired using a Quantum Design Magnetic Property Measurement System. Resistivity and specific heat was measured in a Quantum Design Physical Property Measurement System with a $^3$He option.

\section{Results and discussion}

\subsection{Europium valence}

\begin{figure}
\includegraphics[width=1.0\columnwidth]{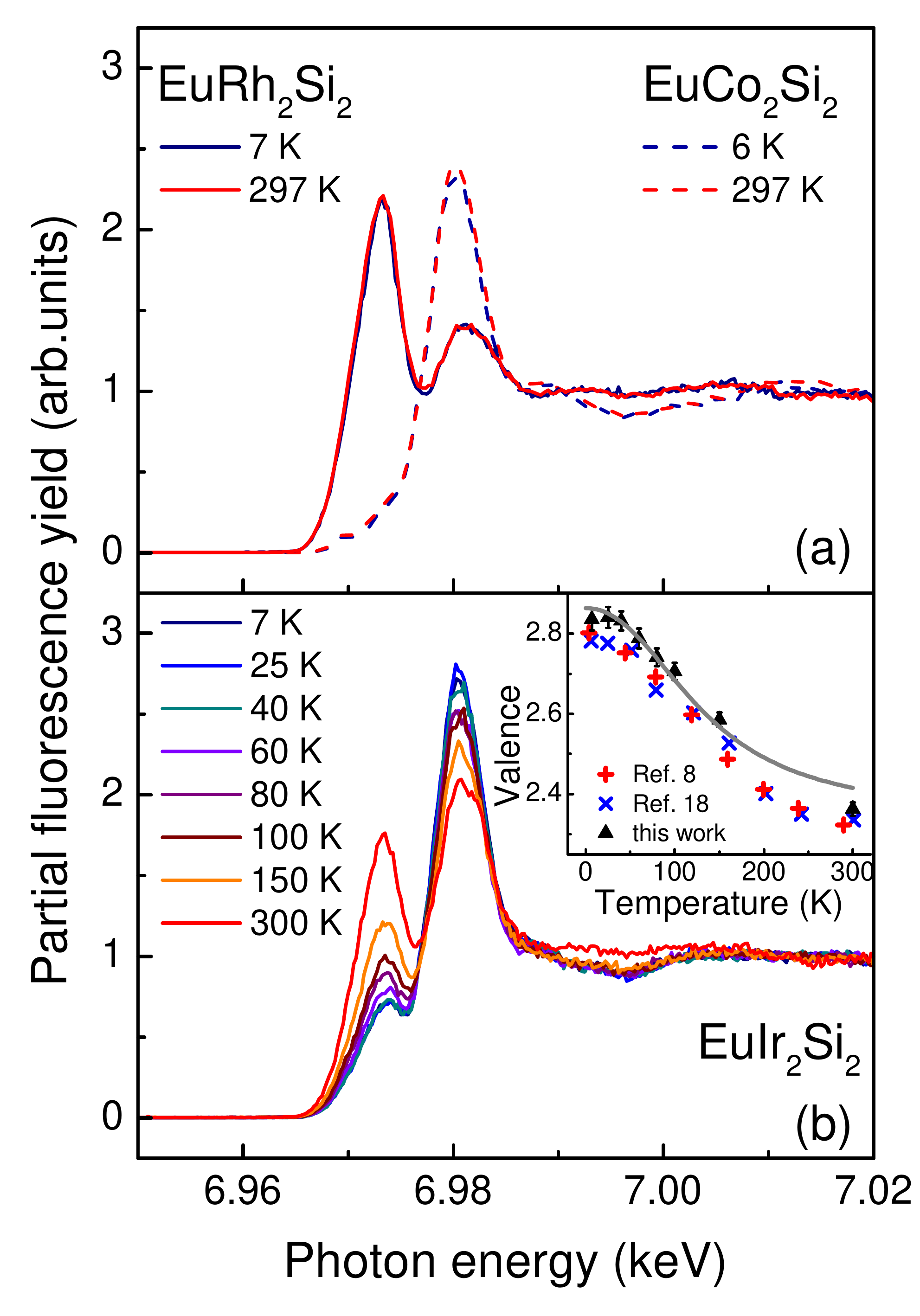}
\caption{\label{absorption} X-ray absorption spectra for (a) \ecs\ and \ers, and (b) \eis. Inset:  Eu valence  in \eis\ as a function of temperature, as determined from the XAS spectra (black triangles, see text) and M\"ossbauer spectroscopy~\cite{Chevalier86,Patil87}. The gray line represents  (\ref{EquationValence}) for $E_{\rm ex}$=(390$\pm$20)\,K, $T_{\rm sf}$=(101$\pm$7)\,K.}
\end{figure}

Figure~\ref{absorption} shows the temperature dependence of X-ray absorption spectra at the Eu L$_3$ edge collected in partial fluorescence yield mode. In the X-ray absorption spectra Eu$^{2+}$ and Eu$^{3+}$ ions give rise to two distinct resonances at about 6.973\,keV and 6.981\,keV, respectively. The shape of Eu L$_3$ XAS spectra is strongly affected by shake-up effects due to the creation of the Eu 2p core-hole.  This is why \ers\ shows a pronounced absorption peak at the characteristic energy of Eu$^{3+}$, although it is well established that Eu is in the divalent state in this compound~\cite{Seiro13}. The relative intensities of the Eu$^{2+}$ and Eu$^{3+}$ absorption lines can therefore not be used directly to determine the Eu valence in the ground state, which lacks the 2p core-hole~\cite{Michels94}. For \ecs\ and \ers\ no difference is detected between spectra taken at low and high temperatures, indicating a stable Eu valence in this temperature range. For \eis, there is a weight shift from the trivalent to the divalent Eu characteristic absorption energies upon increasing temperature. In order to quantify the corresponding valence changes with temperature we have fitted the \eis\ spectra with a linear combination of the \ers\ and \ecs\ spectra using them as a reference for Eu$^{2+}$ and Eu$^{3+}$ absorption, respectively, thus including shake-up effects. The valence obtained in this way is shown as a function of temperature in the inset of figure~\ref{absorption}: it decreases from 2.83 at low temperatures to 2.36 at room temperature. These values are in good agreement with those extracted from the isomer shift in the M\"ossbauer spectra of polycrystalline samples~ \cite{Chevalier86,Patil87}. The non-integer valence at lowest temperatures indicates hybridization between conduction and localized 4f electrons~\cite{Guritanu12}.

Assuming thermal excitation from the $J$=0 ground state of Eu$^{3+}$ to the $J$=7/2 state of Eu$^{2+}$ at an energy $E_{\rm ex}$, and including only the first excited $J$=1 multiplet of Eu$^{3+}$ at  $E_1$ (the next excited multiplets can be neglected because they are at much higher energies), the valence can be expressed as
\begin{equation}\label{EquationValence}
\nu(T) = {\frac {3 (1 + 3  {\rm e}^{-E_1/{\rm k} T^*} ) + 2 ( 8{\rm e}^{-E_{\rm ex}/{\rm k}T^*}) } {1 + 3 {\rm e}^{-E_1/{\rm k}T^*} + 8  {\rm e}^{-E_{\rm ex}/{\rm k}T^*}} },
\end{equation}
where an effective temperature $T^*$=$\sqrt{T^2+T_{\rm sf}^2}$ is introduced, that mimics the effect of hybridization by introducing a phenomenological energy width $T_{\rm sf}$  for all configurations~\cite{Franz80}. For a free Eu$^{3+}$ ion, $E_1$=480\,K, while values in the range 355-505\,K have been reported in Eu compounds~\cite{Seiro13,Samata15}. A fit of the experimentally obtained valence using (\ref{EquationValence}) with $\rm  E_1$=480\,K reproduces the experimental data quite well (see inset of figure~\ref{absorption}(b)), yielding  $E_{\rm ex}$=(390$\pm$20)\,K and $T_{\rm sf}$=(101$\pm$7)\,K. The fit procedure with $E_1$ as a free parameter failed to give physically meaningful parameter values.

\subsection{Structure}

\begin{table*}[ht!]
\caption{Lattice parameters of Eu\textit{M}$_2$Si$_2$ at the highest and lowest measured temperatures. The crystal structure is of the ThCr$_2$Si$_2$-type, space group \textit{I4/mmm}. The uncertainties resulting from the refinement are given in parentheses.}
\begin{tabular} {l r r r r r r}
\br
Compound & \multicolumn{2}{c}{\ecs} & \multicolumn{2}{c}{\ers} & \multicolumn{2}{c}{\eis}\\
\mr
Temperature & 300\,K & 30\,K & 300\,K & 10\,K & 300\,K & 10\,K\\
\mr
$a$ (\AA)	& 3.92218(1) & 3.91258(1) & 4.09085(1) & 4.08341(1) & 4.08353(1) & 4.06569(1) \\
$c$ (\AA)      	&  9.83553(1) & 9.82545(2)  & 10.22471(2) & 10.19238(4) & 10.11054(1) & 10.02233(1)\\
$z$		&  0.3708(1) & 0.3715(1) & 0.3688(2) & 0.3707(2)  & 0.3721(2) & 0.3731(2)\\
$V$ (\AA$^3)$	&151.305(1) & 150.410(1) & 171.111(1) & 169.951(2) & 168.596(1) & 165.668(1)\\
\br
\end{tabular}
\label{struc}
\end{table*}

All three compounds exhibit a tetragonal ThCr$_2$Si$_2$-type structure with space group \textit{I4/mmm} in the whole investigated temperature range. Table~\ref{struc} summarizes the refined structural parameters at the lowest and the highest measured temperatures. Studies on isostructural reference materials such as LaRu$_2$Si$_2$ and LuRh$_2$Si$_2$ have shown that a precise determination of the structural parameters is crucial in order to correctly interpret quantum oscillation results~\cite{Suzuki10,Friedemann13}. While for EuCo$_2$Si$_2$ the parameters at room temperature are consistent with previous works~\cite{Maslankiewicz06}, for EuIr$_2$Si$_2$ the $z$ parameter differs slightly from that reported in the literature~\cite{Rodewald06} and the lattice parameters show strong temperature dependence.  The refined parameter $z$ for \ers~is published here for the first time.

\begin{figure*}
\includegraphics[width=1.0\textwidth]{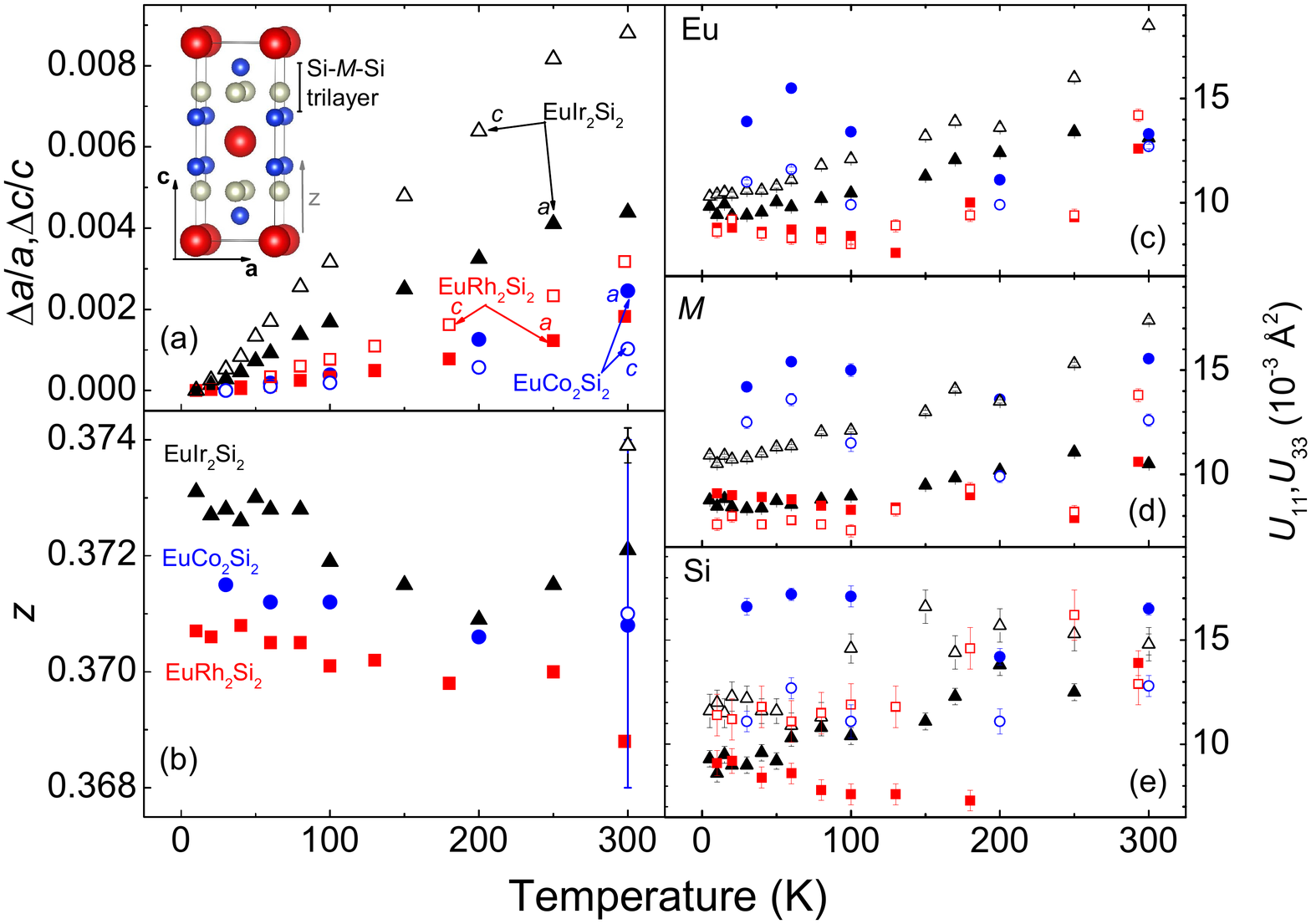}
\caption{\label{thermal expansion} (a) Relative expansion of the lattice parameters $a$ (full symbols) and $c$ (open symbols) with respect to the values at the lowest measured temperature for \ecs~(blue circles), \ers (red squares), and \eis (black triangles). The unit cell is depicted in the inset, with Eu in red, Si in blue and the transition metal in gray. (b) Parameter $z$ determining the position of Si within the unit cell for all three compounds. Previously reported values are plotted for \ecs~\cite{Maslankiewicz06} (blue open circle) and for \eis
~\cite{Rodewald06} (black open triangle). Atomic displacement parameters $U_{11}$ (full symbols) and $U_{33}$ (open symbols) for (c) Eu, (d) transition metal and (e) Si atoms in \eis~(black triangles), \ers~(red squares) and \ecs~(blue circles).  }
\end{figure*}

The temperature evolution of the individual structural parameters is shown in figure~\ref{thermal expansion}. The lattice parameters $a$ and $c$  increase monotonically upon increasing temperature for all compounds. For \eis, however, the lattice expansion is substantially larger than in the stable-valent cases, in particular along the $c$ axis where it reaches almost 1\,\% at room temperature.  The $z$ parameter is weakly temperature dependent, and slightly larger for \eis~than for the stable-valent cases. While thermal expansion is larger along the $c$ axis than along the $a$ axis for the divalent and valence fluctuating compounds, for the trivalent compound thermal expansion along $a$ is larger than along $c$. The latter is in line with results for \textit{R}Cu$_2$Si$_2$ with \textit{R}=trivalent La, Gd and Lu, which exhibit an expansion of about 0.3\,\% along $a$ and 0.1\,\% along $c$ between low and room temperatures~\cite{Neumann85}. In comparison, CaCu$_2$Si$_2$, where Ca is in a divalent configuration, shows between low and room temperatures a  similar lattice expansion along $a$  but twice as large along $c$ than that of the trivalent analogues~\cite{Neumann85}.  Clearly, the rare earth valence affects the bonds along the $c$ axis more strongly than along $a$, making them particularly soft in the case of fluctuating valence. Indeed, while for integer-valent rare earth compounds the bulk modulus scales linearly with $q/V$, where $q$ is the charge of the rare earth ion and $V$ the unit cell volume, analogous compounds where the rare earth valence is  non-integer fall clearly below this trend~\cite{Neumann85}.

Similarly, for \eis, the atomic displacement parameters for Eu and \textit{M} atoms along $c$ ($U_{33}$) are larger and increase more rapidly upon warming than those in the basal plane ($U_{11}$), see figure~\ref{thermal expansion}.  For \ecs, \textit{U}$_{11}$ appears to be larger than \textit{U}$_{33}$ for both Eu and Co atoms, while for \ers~U$_{11}$ and U$_{33}$ are similar.  Due to the smaller cross section of Si compared to that of the heavier atoms, the displacements of Si carry a much larger uncertainty so that no reliable trend can be recognized.

\begin{figure}
\includegraphics[width=1.0\columnwidth]{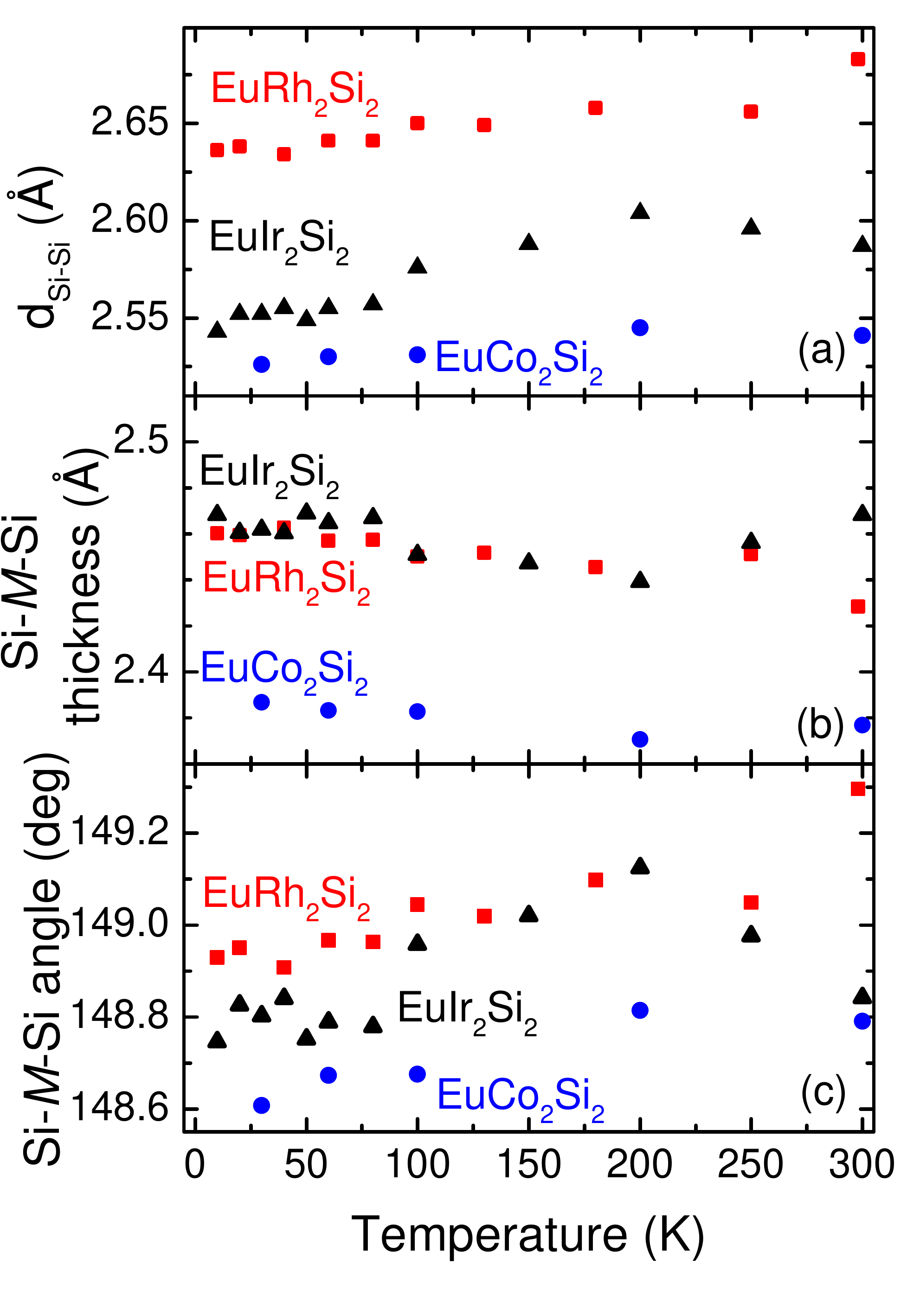}
\caption{\label{bonds} Temperature dependence of (a) the Si-Si distance, (b) the Si-$M$-Si trilayer thickness, and (c) the Si-$M$-Si bond angle.}
\end{figure}

Figure~\ref{bonds}(a) shows that the Si-Si distance depends strongly on the europium valence: it is largest in the divalent compound, smallest in the trivalent compound, and intermediate for the non-integer-valent case. This is in contrast to the thickness of the  Si-\textit{M}-Si  trilayer (defined as the height difference between Si layers across the \textit{M} layer, see inset of figure~\ref{thermal expansion}(a)), which is dominated by the size of the transition metal ion. For both \eis\ and \ers\ the Si-\textit{M}-Si trilayer thickness is esentially identical, in spite of the different valence state of Eu. Similarly, the Si-\textit{M}-Si angles do not show a distinct correlation to the valence state of Eu.  This suggests that the extra conduction electron contributed by Eu$^{3+}$ goes to a band dispersing along the $z$ direction, making the material more three-dimensional. This is reflected, for example,  in a worse cleaving behavior of the \eis\ single crystals compared to \ers\ ones.

\subsection{Thermodynamic and transport properties}

\begin{figure}
\includegraphics[width=1.0\columnwidth]{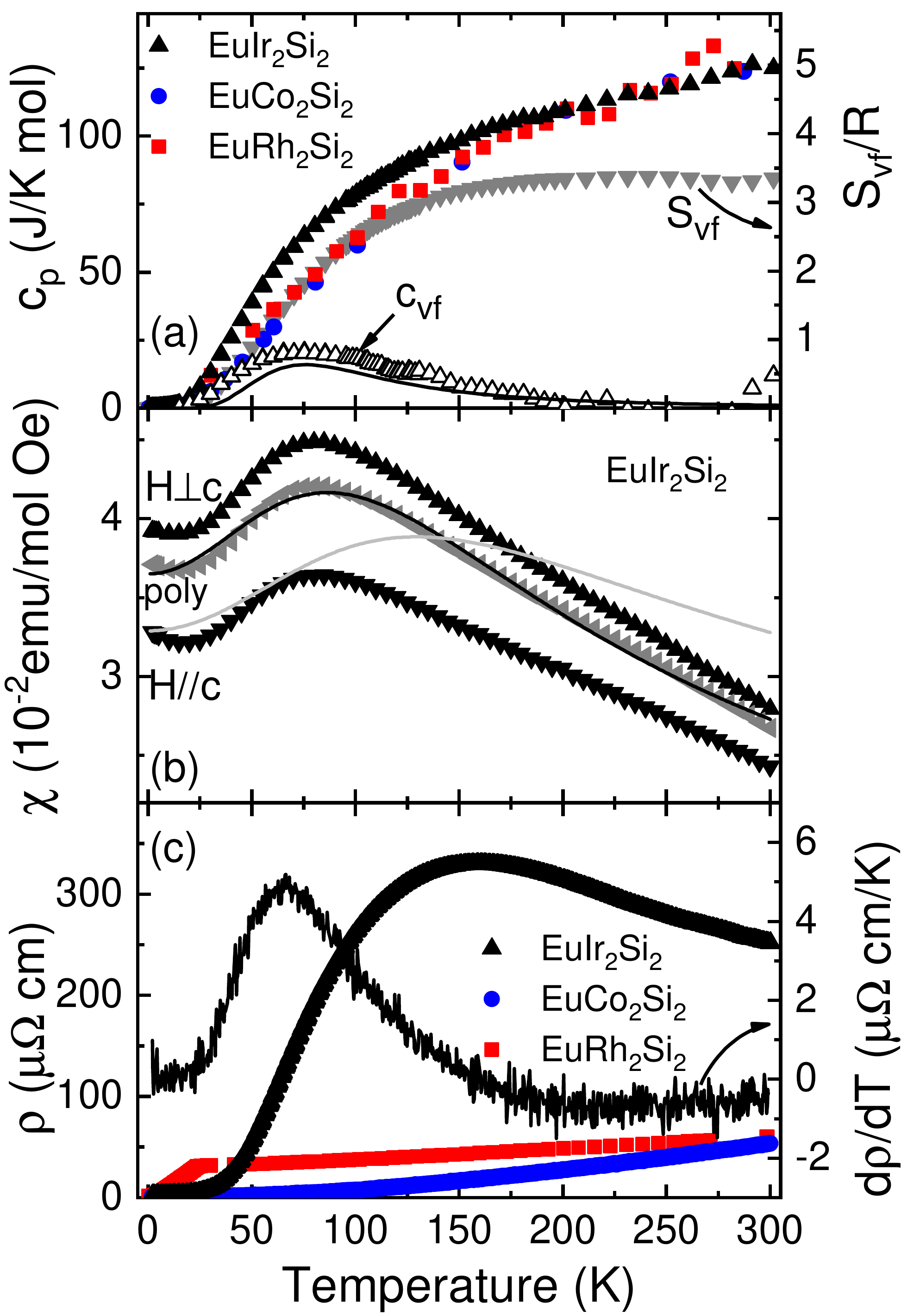}
\caption{\label{specheat}(a) Specific heat of \eis, \ecs~and \ers\ ($T$$>$$T_{\rm N}$).  Excess specific heat due to valence fluctuations in \eis\ is plotted as empty triangles. The solid line represents a Schottky anomaly with $\Delta$=260\,K. Right axis: Entropy due to valence fluctuations in \eis\ in units of the universal gas constant R. (b) Magnetic susceptibility of \eis\  for magnetic fields applied parallel (down triangles) and perpendicular (up triangles) to \textbf{c}, and polycrystalline average (gray triangles). The gray line represents~(\ref{EqS}) for parameters extracted from the valence fit in figure~\ref{absorption}. The black line represents a fit with~(\ref{EqS}) plus a constant susceptibility, yielding $\chi_c$=0.0089\,emu/mol\,Oe, $E_{\rm ex}$=269\,K and $T_{\rm sf}$=84\,K. (c) Resistivity for current along the basal plane (left axis), and its derivative as a function of temperature for \eis\ (black line, right axis).}
\end{figure}

Specific heat as a function of temperature is plotted in figure~\ref{specheat}(a) for all three compounds. At low temperatures, the specific heat of \eis~behaves as expected for a Fermi liquid, with an electronic contribution linear in temperature given by the Sommerfeld coefficient $\gamma$=33\,mJ/K$^2$mol, and a phonon contribution characterized by a Debye temperature $\Theta_{\rm D}$=270\,K ~\cite{Seiro11}. This value of $\gamma$ is higher than for the stable-valent compounds (9.6\,mJ/K$^2$mol for \ecs~\cite{Seiro13}, $\sim$25\,mJ/K$^2$mol for \ers~\cite{Seiro11}), indicating a moderate renormalization of the effective mass in agreement with optical spectroscopy studies~\cite{Guritanu12}. The Debye temperature is much smaller than for the stable-valent compounds (388\,K for \ers~\cite{Seiro13JPCM},  332 K  for \ecs~\cite{Seiro13}). Though a smaller $\Theta_{\rm D}$ is expected due to the higher mass of iridium atoms, the large thermal expansion in figure~\ref{thermal expansion}(a) suggests that the low value of $\Theta_{\rm D}$ for \eis\ is related to the phonon softening induced by valence fluctuations. 

\eis~ exhibits a clear excess specific heat with respect to the stable-valent  compounds that extends over a broad temperature range, roughly 20 to 200\,K.  In an attempt to quantify this excess specific heat, we subtracted from the data for \eis~an average of the specific heat of \ers~and \ecs~weighted by the actual valence of Eu in \eis.  Notice that this procedure only yields an approximate estimate, since the different atomic masses of Ir, Rh and Co are not considered. The contribution of valence fluctuations to the specific heat peaks at roughly 75\,K and is basically negligible below 20\,K and above 200\,K. This is consistent with the temperature evolution of the valence in figure~\ref{absorption}, which shows only weak changes at the lowest and highest temperatures. Predictions for the specific heat due to valence fluctuations within a realistic model  are not available, but calculations for the spinless Falicov-Kimball model suggest that for a large energy difference between the localized f state and the Fermi level, specific heat shows a Schottky-like behavior~\cite{Farkasovsky96}.  In a simple two-level Schottky model with non-degenerate ground state and an excited state at an energy $\Delta$ with a degeneracy of 8 (that of a  Eu$^{2+}$ configuration), a maximum at 75\,K corresponds to $\Delta\sim260$\,K, a value significantly lower than that of $E_{\rm ex}$ obtained from the thermal evolution of the valence in figure~\ref{absorption}. The temperature dependence of the Schottky specific heat for this value of $\Delta$, plotted as a solid line in figure~\ref{specheat}(a), roughly follows that of the excess specific heat due to valence fluctuations in \eis, albeit with somewhat smaller values than the experimental data. This difference possibly reflects the lower Debye temperature of \eis~with respect to the stable valent compounds, that would be consistent with the softening observed in figure~\ref{thermal expansion}.  Accordingly, the estimated entropy associated to the valence fluctuations saturates at high temperatures at values beyond the ${\rm R} \ln 9$ expected for the Schottky model.

The peak in the specific heat contributed by valence fluctuations correlates with a peak in the magnetic susceptibility of \eis,  observed both for fields applied parallel and perpendicular to the tetragonal $c$ axis, see figure~\ref{specheat}(b). Such a broad maximum is typical of fluctuating valence compounds. At higher temperatures, the susceptibility decreases upon warming but no clear Curie-Weiss-type behavior is recovered up to room temperature. At low temperatures, the susceptibility saturates at a residual value of 0.03-0.04 emu/mol Oe that is relatively high for a metal and can be at least partly ascribed to a Van Vleck contribution involving the higher $J$ multiplets of Eu$^{3+}$. A small anisotropy is observed, which decreases on increasing temperature. This anisotropy is possibly related to the crystal-field splitting of the excited multiplets of Eu$^{3+}$, and has also been reported for the trivalent Eu compound \ecs~\cite{Seiro13}. The fact that the susceptibility for fields along $c$ is smaller than for fields applied in the basal plane indicates that the $J_z$=0 levels lie higher in energy within each $J$ multiplet~\cite{Tovar89}. Considering only the $J$=1 excited multiplet, if the observed susceptibility at low temperatures was purely the Van Vleck contribution of Eu$^{3+}$, this would correspond to a splitting of 16\,K between $J_z$=0 and $J_z$=$\pm$1 states. However, a pure Van Vleck-like contribution implies for the average residual susceptibility $\chi_0\sim$0.036\,emu/mol an energy for the first excited $J$=1 multiplet of about $E_1\sim$80\,K, an unrealistically low value. Since the valence of Eu at low temperatures remains far from 3, and renormalization effects are observed e.g. in the Sommerfeld coefficient, a more realistic comparison can be made with a weighted average of divalent (Curie) and trivalent (Van Vleck) contributions, in the spirit of~(\ref{EquationValence}), where the weight is given by the valence~\cite{Franz80}: 

\begin{equation}
\label{EqS}
\chi(T) = 8{\rm N}\mu_{\rm B}^2 \frac{ \frac{21{\rm e} ^{-E_{\rm ex}/{\rm k}T^*}}{{\rm k} T} + \frac{1}{E_1} + {\rm e}^{-E_1/{\rm k}T^*} (\frac{1}{{\rm k} T }-\frac{1}{ E_1})}{1 + 3{\rm e}^{-E_1/{\rm k}T^*}+ 8{\rm e} ^{-E_{\rm ex}/kT^*}}
\end{equation}

\noindent and a renormalized temperature $T^*=\sqrt{T^2+T_{\rm sf}^2}$ is considered, as in~(\ref{EquationValence}). However, substituting in~(\ref{EqS}) the values for $E_1$, $E_{\rm ex}$ and $T_{\rm sf}$ extracted from the fit of~(\ref{EquationValence}) to the valence gives a poor agreement with the experimental susceptibility data, even after addition of a constant susceptibility term $\chi_c$, see gray line in figure~\ref{specheat}(b). A better agreement is obtained using $ E_1$=480\,K, which is represented by the black line, yielding $\chi_c$=(0.0089$\pm$0.0001)\,emu/mol\,Oe, $E_{\rm ex}$=(269$\pm$1)\,K and $T_{\rm sf}$=(84$\pm$1)\,K. The latter parameters are quite different to those obtained from the analysis of the valence. 

The dramatic effect of valence fluctuations in transport properties is shown in figure~\ref{specheat}(c). While at lowest temperatures the resistivity of \eis\ and that of stable-valent \ers\ and \ecs\ are of similar magnitude, the resistivity of \eis\ increases over two orders of magnitude upon warming in the temperature range where strong valence fluctuations are observed in the specific heat, reaching a maximum at around 160\,K.  Upon further warming, resistivity decreases on increasing temperature and stagnates towards room temperature. Several theoretical works~\cite{Zipper87,Capellmann88,Zlatic2001} indicate the presence of a resistivity maximum as a consequence of conduction electrons scattering off valence fluctuations, although the shape of the resistivity peak depends on the particular model and parameter range. In a simple picture, the lifetime of conduction electrons decreases on increasing temperature, becoming of the order of the characteristic valence fluctuation time  and leading to a large rise in resistivity. Resistivity then saturates as the scattering becomes fully incoherent at higher temperatures. The total resistivity increase due to valence fluctuations $\Delta\rho$ has been proposed to scale with $\Delta\nu/T_{\rm sf}$, where $\Delta\nu$ is the percent deviation from integer valence at high temperatures, with $\Delta\rho (\Delta\nu/T_{\rm sf})^{-1}$$\sim$ 750-900\,$\mu\Omega$cmK for several Ce- and Yb-based valence fluctuating compounds~\cite{Zipper87}. Estimating the phonon contribution from $\rho$(300\,K)-$\rho$($T_{\rm N}$) for \ers\, yields for \eis\ $\Delta\rho (\Delta\nu/T_{\rm sf})^{-1}$$\sim$1000-1250\,$\mu\Omega$cmK. The origin of the large resistivity increase is evident when comparing the derivative of resistivity with respect to temperature with the specific heat contribution from valence fluctuations: both exhibit roughly the same shape, with a well-defined maximum at $\sim$65\,K. The correspondence of resistivity change and specific heat has been established for magnetic systems~\cite{Campoy06} and supports the observation of an excess specific heat due to valence fluctuations in \eis.  For stable-valent compounds, resistivity shows a much slower monotonous increase on increasing temperature that can be well described by phonon scattering~\cite{Seiro13, Seiro13JPCM}. 

At low temperatures, the resistivity of \eis\ follows $\rho$=$\rho_0$+$A T^2$, with $\rho_0$=(2.67$\pm$0.03)\,$\mu\Omega$cm and $A$=(0.0026$\pm$0.0005)\,$\mu\Omega$/cmK$^2$. This dependence, as well as a specific heat varying as c=$\gamma T$ with a moderately high $\gamma$ value, are characteristic of a correlated metal. The ratio $A/\gamma^2$, known as the Kadowaki-Woods ratio~\cite{Kadowaki86}, shows a universal value within certain groups of materials, e.g. 0.4\,$\mu\Omega$\,cm\,mol$^2$K$^2$/J$^2$ for transition metal elements ~\cite{Rice68} and  10(0.36)\,$\mu \Omega$\,cm\,mol$^2$K$^2$/J$^2$ for Kondo systems with twofold (eightfold) degeneracy~\cite{Tsuji03}.  In the case of \eis, $A/\gamma^2$$\sim$2.4$\,\mu\Omega$\,cm\,mol$^2$K$^2$/J$^2$, is close to the values observed for intermediate-valent EuCu$_2$(Si$_{0.7}$Ge$_{0.3}$)$_2$ and EuNi$_2$P$_2$ (2.25 and 2.7 $\mu\Omega$\,cm\,mol$^2$K$^2$/J$^2$, respectively)~\cite{Hossain04,Hiranaka13}. A further indicator of correlations is the Wilson ratio $R_{\rm W}=\pi^2k_{\rm B}^2\chi_0/ (\mu_{\rm eff}^2\gamma)$, where $\chi_0$ is the Pauli susceptibility and $\mu_{\rm eff}$ an effective magnetic moment.  In absence of correlations (as in a Fermi gas) $R_{\rm W}$=1, while for the Kondo impurity model $R_{\rm W}$=2 in the two-fold degenerate case~\cite{Wilson75}.  Taking $\mu_{\rm eff}=7.94$, as for Eu$^{2+}$, results in $R_{\rm W}$=3.9.  This relatively large value compared to 1.15 obtained for EuCu$_2$(Si$_{0.7}$Ge$_{0.3}$)$_2$~\cite{Hossain04}, and $\sim$1.4 for EuNi$_2$P$_2$, is probably related to the fact that the low-temperature valence of Eu in \eis\ is closer to 3 and the system is further away from the magnetic critical point. As a consequence,  the susceptibility of \eis\ at low temperatures is dominated by the Van Vleck rather than by the Pauli contribution. This is in line with $R_{\rm W}$=3.1 for \ecs\, where Eu is trivalent and correlations are weaker ($\gamma$=9.6\,mJ/K$^2$mol)~\cite{Seiro13}.

 \section{Conclusions}

In conclusion,  two different aspects are reflected in the properties of \eis. On one hand, the fluctuating valence as a function of temperature manifests in a large thermal expansion, broad peaks in the excess specific heat, magnetic susceptibility and a dramatic increase in resistivity. On the other hand, the intermediate valence at low temperatures is accompanied by Fermi-liquid  behavior with a moderate electron mass renormalization, as observed in specific heat, susceptibility and resistivity. The interconfiguration fluctuation model, a simple phenomenological model combining contributions from  Eu$^{2+}$ and  Eu$^{3+}$ weighted by the Bolzmann factor, and that mimics hybridization effects through a renormalized temperature, reproduces the behavior of valence and magnetic susceptibility independently, but is unable to do so with a single set of parameters, thus showing the limitations of the model for quantitative analysis of the experimental data.

\section*{Acknowledgements}

Dr. Yves Watier is acknowledged for the assistance during the synchrotron powder diffraction experiment at beamline ID22 at ESRF (proposal HS1471).

\end{document}